\documentclass[prb,twocolumn,amsmath,aps,pdffig,showpacs,superscriptaddress]{revtex4}
\usepackage{graphicx}
\usepackage{dcolumn}
\usepackage{bm}
\usepackage{hyperref}
\usepackage{soul}
\usepackage{color}
\usepackage{ulem}

\begin{document}

\title{Pseudospin polarized quantum transport in monolayer graphene}

\author{Leyla Majidi and Malek Zareyan}

\affiliation{Institute for Advanced Studies in Basic Sciences
(IASBS), P. O. Box 45195-1159, Zanjan 45195, Iran}
%
\begin{abstract}
Monolayer graphene with an energy gap presents a pseudospin
symmetry broken ferromagnet with a perpendicular pseudomagnetization
whose direction is switched by altering the type of
doping between n and p. We demonstrate an electrical current
switching effect in pseudospin version of a spin valve in which
two pseudoferromagnetic regions are contacted through a normal
graphene region. The proposed structure exhibits a pseudomagnetoresistance,
defined as the relative difference of
resistances of parallel and antiparallel alignments of the pseudomagnetizations,
 which can be tuned to unity. This perfect
pseudomagnetic switching is found to show a strong robustness with
respect to increasing of the contact length, the effect which we
explain in terms of an unusually long range penetration of an
equilibrium pseudospin polarization into the normal region by
proximity to a pseudoferromagnet. Our results reveals the
potential of gapped graphene for realization of pseudospin-based
nanoelectronics.

\end{abstract}
%

\maketitle


\section{\label{sec:intro}Introduction}

Graphene, the two dimensional layer of the carbon atoms with honeycomb lattice structure, has attracted a great deal of attention as
a new promising material for nanoelectronics, since its experimental realization a few years ago\cite{Novoselov04,Novoselov05,Zhang05}.
Most of the peculiar properties of graphene is the result of its massless Dirac spectrum of the low-lying electron-hole excitations, which
in addition to the regular spin appear to come endowed with the two quantum degrees of freedom, the so called pseudospin and valley. The
pseudospin represents the sublattice degree of freedom of the graphene's honeycomb structure, and the valley defines the corresponding
degree of freedom in the reciprocal lattice\cite{Novoselov05,Wallace74,Slonczewski58,Haldane88,Castro09}. The effect of these additional
quantum numbers has already been exploited by anomalous features of several quantum transport phenomena in graphene, including quantum Hall
effect\cite{Novoselov05,Zhang05,Gusynin05,Du09}, conductance quantization\cite{Peres06}, Klein tunneling\cite{Katsnelson06,Young09,Stander09}
and quantum shot noise\cite{Tworzydo06,Danneau08,DiCarlo08}.
\par
Interestingly, the pseudospin and valley degrees of freedom in graphene have been proposed separately to be used for controlling the electronic
devices in the same way as the electron spin is used in spintronic and quantum computing. Rycerz {\it et al.}\cite{Rycerz07,Xiao07,Akhmerov08}
demonstrated an electrostatically controlled valley filter effect in graphene nanoribbons with zigzag edge which can be used for realizing valley
valve structures in valleytronics (valley-based electronics) applications. On the other hand a pseudospin-based version of
a spin valve has been proposed in bilayer graphene, where the pseudospin is determined by the relative amplitude of the wave function on the two
layers\cite{Jose09}. A bilayer pseudospin valve consists of two connected neighboring regions whose pseudospin polarizations can be tuned by application of gate voltages \cite{xia10}.
\par
In this paper, we study the possibility of realizing
pseudospintronics in monolayer graphene within the scattering
formalism. The possibility of an interaction driven spontaneous
breaking of the pseudospin symmetry, which can lead to the
realization of pseudomagnetic states in monolayer and bilayer
graphene, has been studied recently\cite{Min08}. Here, we
demonstrate that the monolayer graphene with a gap in its
electronic spectrum and an appropriate doping presents a
pseudospin symmetry broken ferromagnet, with a finite pseudospin
magnetization oriented vertically to the graphene plane. The
magnitude of the pseudomagnetization (PM) depends on the chemical
potential and its direction is switched by changing the type of
doping (electron n or hole p).
\par
Based on the above observation, we propose a nonmagnetic
pseudospin valve which consists of two pseudoferromagnetic (PF) regions
 separated by a non-pseudomagnetizaed normal (N) graphene of
length L (shown schematically in Fig. \ref{Fig:1}b). The PM
direction in each region can be tuned independently by means of electrical gates,
which allows for switching between parallel and antiparallel configurations. We
find that the proposed PF/N/PF spin valve exhibits a
relative difference of the electrical resistance in
the two parallel and antiparallel states of the pseudomagnetizations, which can be remarkably large in analogy to the giant
magnetoresistance (GMR) in magnetic multilayers\cite{Baibich88}.
When the chemical potential of the system is tuned to be close to
the energy gap ($\mu\simeq\Delta$), the pseudospin valve effect
can be perfect PMR takes the value 1 for appropriate choices of the length
$L$. More importantly, we show that the perfect pseudospin valve
effect can be reached even in higher chemical potentials
$\mu\gg\Delta$ by applying an appropriate bias voltage. We further
demonstrate the unusual proximity effect at
pseudoferromagnet-normal junctions (PF/N, PF/N/PF). We find that
an equilibrium pseudospin polarization is induced into the N region
with a direction which is precessing around the axis normal to the
PF/N interface, and an amplitude which decays slowly in the
distances of few Fermi wave length $\lambda_F$ from the interface.
This is in clear contrast to the induced magnetization in ordinary
ferromagnet-normal metal junctions, which decays exponentially
within $\lambda_F$\cite{Zutic04}.
\par
This paper is organized as follows. In Sec.\ \ref{sec:level1}, we
introduce pseudoferromagnets (PF) and use them to study the
pseudospin valve effect in monolayer graphene PF/N/PF junction.
Sec.\ \ref{sec:level2} is devoted to the investigation of the
proximity effect in graphene PF/N junctions. Finally, we present
the conclusion in Sec.\ \ref{sec:level3}.


\section{\label{sec:level1}Pseudospin valve}
\begin{figure}
\begin{center}
\includegraphics[width=3.5in]{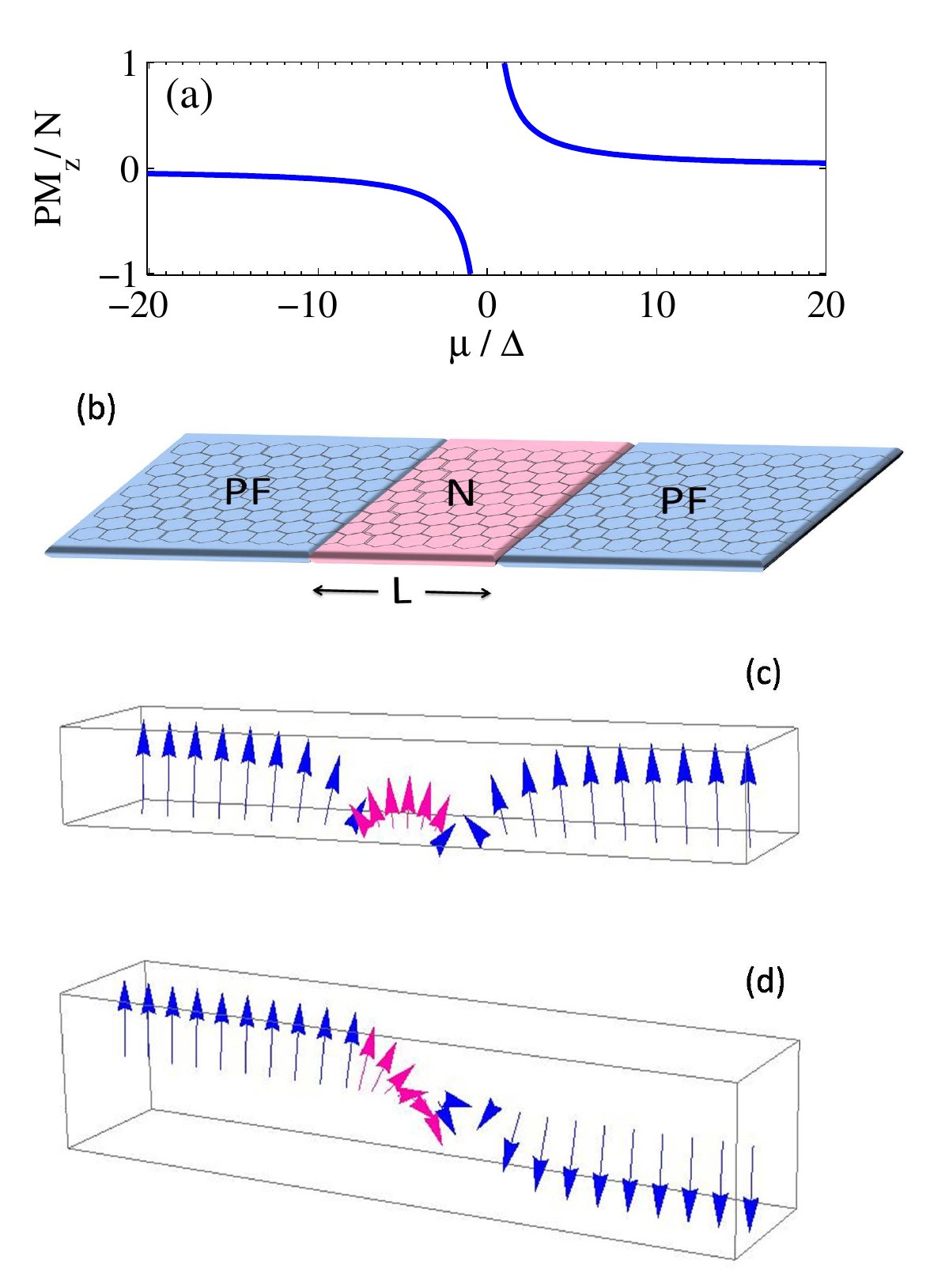}
\end{center}
\caption{\label{Fig:1}(Color online) (a) Vertical
pseudomagnetization per electron $PM_z/N$ of the gapped graphene layer
versus chemical potential $\mu$ ( $\mu$ is scaled to the energy gap $\Delta$). (b) Schematic illustration of the proposed
pseudospin valve in monolayer graphene: The left and right regions
are pseudoferromagnets (PF) and the intermediate region is a normal
graphene (N) without a band gap. (c-d) Profile of pseudomagnetization vector $\bm{PM}$
inside the two PFs (blue) and the N region of
length $L=\lambda_F$ (pink) for two configurations of (c) parallel
and (d) anti-parallel, when $\mu\simeq\Delta$.}
\end{figure}
The pseudospin valve structure we study is shown schematically in
Fig.~\ref{Fig:1}b. It consists of two PF regions
with tunable direction of PM which are
connected through a normal (non-pseudomagnetized) layer of length
$L$. Such a structure can be realized in a graphene sheet with a
sizable gap in its electronic band structure. There are several
methods to open an energy gap in the band structure of graphene. A
scenario is placing graphene on top of an appropriate substrate
which breaks the graphene sublattice symmetry and generates a
Dirac mass for charge carriers. The band gap opening is observed
in epitaxially grown graphene on a SiC
substrate\cite{Zhou07,Varchon07} and a hexagonal boron nitride
crystal\cite{Giovannetti07}, and also on metallic surfaces with a
boron nitride buffer layer\cite{Lu07}. The energy band gap
engineering can be also achieved through doping the graphene with
several molecules such as $CrO_3$, $NH_3$, $H_2O$
\cite{Zanella08,Ribeiro08}. A gapped graphene with a chemical
potential $\mu$ close to its energy gap $\Delta$ acquires a
PM of sizable magnitude and an orientation which
switches with changing the sign of the chemical potential (see
Fig.~\ref{Fig:1}a). A highly doped gapped graphene has a
vanishingly small PM and behaves as a
non-pseudomagnetized (normal) graphene.
\par
To study quantum transport in the pseudospin valve structure
within the scattering formalism, we first construct the
quasiparticle wave functions that participate in the scattering
processes. We adopt Dirac equation of the form
\begin{equation}
\label{DiracFull} H\psi=(\varepsilon+\mu)\psi,
\end{equation}
where in the presence of an energy gap
$H=v_{F}(\bm{\sigma}\mathbf{}.\bm{p})+\Delta \sigma_{z}$ is the
two-dimensional Dirac Hamiltonian with $\bm{p}=-i\hbar\bm{\nabla}$
the momentum operator in the x-y plane ($v_F=10^6$ m/s represents
the Fermi velocity) and $\bm{\sigma}=(\sigma_x,\sigma_y,\sigma_z)$
the vector of the Pauli matrices operating in the space of two
sublattices of the honeycomb lattice\cite{Divincenzo84,Ando05}.
The two-dimensional spinor has the form $\psi=(\psi_A,\psi_B)$,
where the two components give the amplitude of the wave function
on the two sublattices and $\varepsilon$ is the
quasiparticle energy.
\par
For a uniform gapped graphene region the solutions of the Dirac
equation Eq. (\ref{DiracFull}) are two states of the form
\begin{equation}
\label{psi_c}
\psi_{c}^{e\pm}=e^{\pm ik_cx} e^{iqy}
\left(
\begin{array}{c}
e^{\mp i\alpha_c/2}\\
\pm e^{-\phi_c} e^{\pm i\alpha_c/2}
\end{array}
\right),
\end{equation}
for conduction band electrons of n-doped graphene and
\begin{equation}
\label{psi_v}
\psi_v^{e\pm}=e^{\mp ik_vx} e^{iqy}
\left(
\begin{array}{c}
e^{\pm i\alpha_v/2}\\
\pm e^{\phi_v} e^{\mp i\alpha_v/2}
\end{array}
\right),
\end{equation}
for valance-band electrons of
p-doped graphene, at a given energy $\varepsilon$ and transverse wave
vector q with the energy-momentum relation $\varepsilon_{c,
v}=\pm[-\mu+\sqrt{\Delta^2+(\hbar v|{\bm{k}}_{c, v}|)^2}]$. Here
$\alpha_{c, v}=\arcsin({\hbar vq}/{\sqrt{(\varepsilon\pm\mu)^2-\Delta^2}})$ is the angle of
propagation of electron which has longitudinal wave vector $k_{c,v}=(\hbar v_F)^{-1}\sqrt{(\varepsilon\pm\mu)^2-\Delta^2}\cos\alpha$ and
$\phi_{c, v}=
arcsinh{(\Delta/\sqrt{(\varepsilon\pm\mu)^2-\Delta^2})}$. The two propagation directions of electron along the x axis are denoted by $\pm$ in $\psi_{c,v}^{e\pm}$.
\par
The pseudospin of such states for conduction (valance) band
electrons of n- (p-)doped graphene is obtained as
\begin{equation}
\label{pseudospin}
<\bm{\sigma}(\bm{k})>_{\psi_{c, v}}=\sqrt{1-(\frac{\Delta}{\varepsilon\pm\mu})^2}\ \hat{\bm{k}}_{c,v}+\frac{\Delta}{\varepsilon\pm\mu}\ \hat{\bm{k}}_{\bot},
\end{equation}
where $\hat{\bm{k}}_{\bot}$ is the unit vector normal to the
electronic wave vector $\bm{k}_{c,v}$.
\par
As can be seen from the above equation, the
existence of a band gap makes the pseudospin
to have a component perpendicular to the
electronic momentum in the plane of the graphene sheet. The in-plane and out of plane
components of the pseudospin depend on $(\varepsilon+\mu)/\Delta$,
which can be tuned to unity to make the pseudospin vector to be
oriented perpendicular to the sheet. Increasing
$(\varepsilon+\mu)/\Delta$ leads to the decrease of the out of
plane component such that it goes to zero when
$\varepsilon+\mu\gg\Delta$.
\par
The total PM of the gapped graphene is calculated
by summing the expression (\ref{pseudospin}) over all the wave
vectors $\bm{k}=(k,q)$,
\begin{equation}
\label{pseudomagnetization}
\bm{PM}_{n,p}=\sum_{\bm{k}}<\bm{\sigma}(\bm{k})>_{\psi_{c,v}},
\end{equation}
from which we find that PM only has an out of plane component which depends on
$(\varepsilon+\mu)/\Delta$. Fig. \ref{Fig:1}a shows the behavior
of the out of plane component of PM per electron $PM_z/N$ as a
function of $\mu/\Delta$ at zero temperature
($T=0$). It is seen that
for $\mu\simeq\Delta$, $PM_z/N$ takes its maximum value $PM_z/N=1$
, while increasing $\mu/\Delta$ leads to the decrease of $PM_z$ such
that it goes to zero for highly doped gapped graphene
($\mu\gg\Delta$). For a given value of $|\mu|$, the orientation of
PM changes by changing the sign of $\mu/\Delta$. This implies that
the PM vectors of gapped graphene regions with different type (n
or p) of dopings are oriented antiparallel versus each other.
\par
\begin{figure}
\begin{center}
\includegraphics[width=3in]{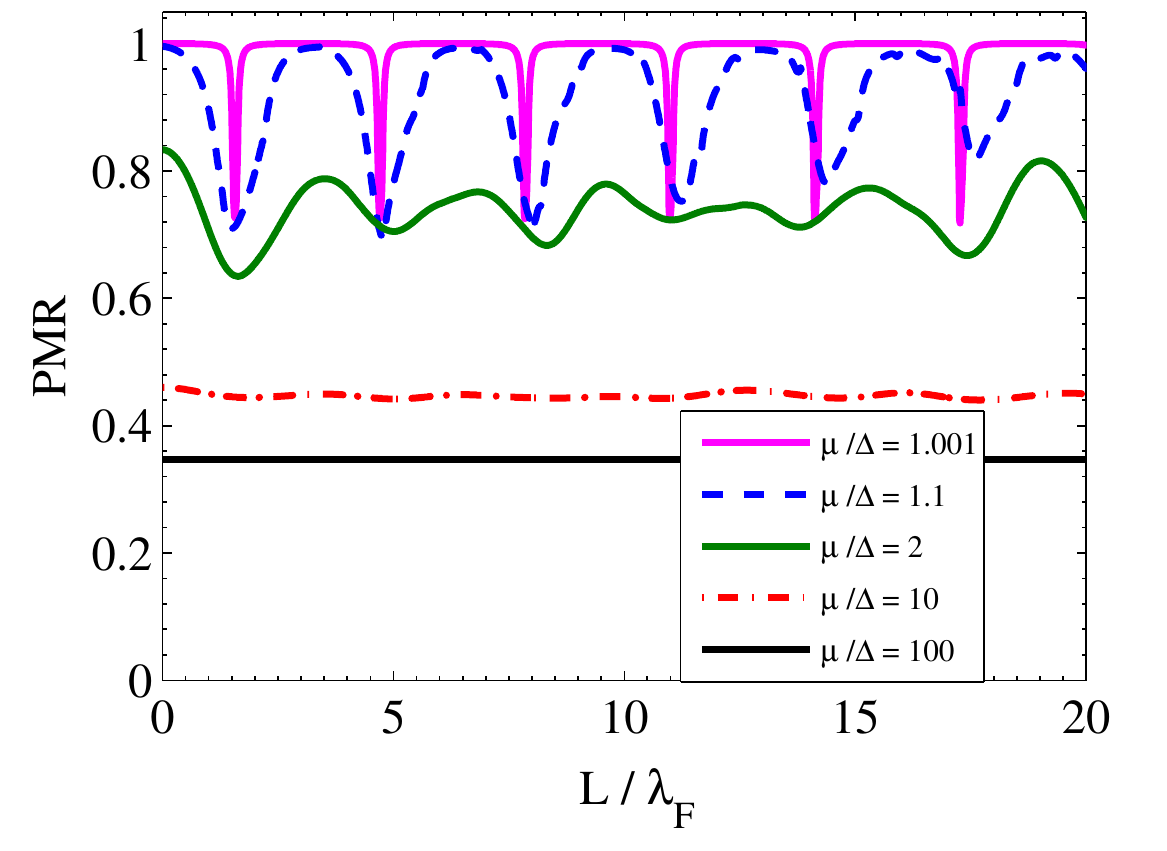}
\end{center}
\caption{\label{Fig:2}(Color online) Pseudomagnetoresistance (PMR) of
the pseudospin valve versus the length of the N region
($L/\lambda_F$) for different values of $\mu/\Delta$. }
\end{figure}
We note that the pseudospin polarization of a gapped graphene
corresponds to a difference in the electronic charge densities of
the two triangular sublattices, which in turn produces an
in-plane electrical polarization of the density $\bm{P}_{n,p}={\sum_{\bm{k}}}\bm{p}_{n,p}$.
An electron with a wave vector $\bm{k}$ contributes a dipole moment which is calculated from the relation
\begin{eqnarray}
\label{dipole}
\bm{p}_{n,p}&=&\pm(1-|\frac{\psi_B}{\psi_A}|_{c,v}^{\pm2})\ e\ \bm{a}_{AB}\nonumber\\
&=&\pm(1-e^{-2\phi_{c,v}})\ e\ \bm{a}_{AB},
\end{eqnarray}
 where $\bm{a}_{AB}$ is the lattice vector oriented from A to
B site. We see that the electric dipole
moments are oriented from A to B (B to A) sites for n- (p-)doped
gapped graphene. This correspondence between PM vector
and the density of the in-plane electrical polarization can be used
for an experimental measuring of PM.
\par
With the above found behavior of PM, the realization of the pseudospin valve of Fig. \ref{Fig:1}b seems quite feasible. The
configuration of the pseudospin valve can be changed from
parallel to antiparallel by fixing the type of doping of one
region and changing the type of the doping in the other region.
The size of the pseudospin valve effect is determined by the
extent in which the conduction of the anti-parallel configuration
is suppressed (similar to the spin valve effect). The
pseudomagnetoresistance of a pseudospin valve is
defined as
\begin{equation}
\label{PMR}
PMR=\frac{R_{AP}-R_P}{R_{AP}}=\frac{G_{P}-G_{AP}}{G_{P}},
\end{equation}
where $R_{P(AP)}$ is the resistance of the parallel
(anti-parallel) configuration and $G_{P(AP)}$ is the corresponding
conductances, which can be calculated from the Landauer
formula\cite{landauer88}
\begin{equation}
\label{G}
G_{P(AP)}=g_0\int |t_{P(AP)}|^2 \cos\alpha\ d\alpha,
\end{equation}
where $T_{P(AP)}=|t_{P(AP)}|^2$ is the transmission probability of
electrons through the pseudospin valve in parallel
(anti-parallel) configuration and $g_0= 4 e^2/h$ is the quantum of
conductance.
\par
We have calculated the amplitudes of the transmission $t_{P(AP)}$
by matching the wave functions of three regions of the left PF, N
region and the right PF (signed by 1,2, and 3, respectively) at
the two interfaces, $x=0$ and $x=L$.  The solutions of Eq. (\ref{DiracFull}) in
the three regions for parallel (P) and antiparallel (AP)
configurations are as follow
\begin{eqnarray}
\label{t_p}
&&\hspace{-0.5cm}
\psi_1=\psi_c^{e+}+r\ \psi_c^{e-},\nonumber\\
&&\hspace{-0.5cm}
\psi_2=a\ {\psi'_c}^{e+}+b\ {\psi'_c}^{e-},\nonumber\\
&&\hspace{-0.5cm}
\psi_{3,P(AP)}=t_{P(AP)}\ \psi_{c(v)}^{e+}.
\end{eqnarray}
$\psi_{c(v)}^{(')e\pm}$ are the wave
functions of Dirac equation for incoming and outgoing electrons of
n- (p-)doped graphene sheet with (without) a gap; $r$ is the reflection coefficient in
the left PF  and $a$ and $b$ are the coefficients of the states $\psi_{c}^{'e\pm}$.
Matching these solutions at the interfaces, we obtain the transmission amplitudes as
\begin{equation}
\label{t_P}
t_P=\frac{A\ e^{-ik_3L}}{C+D+e^{i\alpha_3-\phi_3}\
(B+(e^{2ik_2L}-1)e^{-i\alpha_2})},
\end{equation}
\begin{equation}
\label{t_AP}
t_{AP}=\frac{A\
e^{ik_3L}}{e^{\phi_3}(B+e^{i\alpha_2}(1-e^{2ik_2L}))+C+De^{i\alpha_3}},
\end{equation}
where $A=4\cos\alpha_1 \cos\alpha_2\
exp{(i[\frac{\alpha_1+\alpha_3}{2}+\alpha_2+k_2L])}$,
$B=exp{(i\alpha_1 +i\phi_1)}\ [exp{(2i\alpha_2)}+exp{(2ik_2L)}]$,
$C=[1-exp{(2ik_2L)}]\ exp{(i[\alpha_1+\alpha_2]+\phi_1)}$ and
$D=1+exp{(2i\alpha_2+2ik_2L)}$.
\par
Finally from the calculated expressions of $G_P$ and $G_{AP}$ via
Eq. (\ref{G}) and using Eq. (\ref{PMR}),
we obtain PMR as a function of the chemical potential and the
length of the N region. Fig. \ref{Fig:2} shows dependence of
PMR on $L/\lambda_{F}$($\lambda_F=\hbar v_F/\mu$) at $T=0$ and zero bias
voltage $V=0$ and for different values of $\mu/\Delta$. We have
taken $\mu_1=\mu_2=|\mu_3|=\mu$. We observe that the pseudospin
valve effect can be perfect ($PMR=1$) for $\mu\simeq\Delta$. For these
values of $\mu$, PMR shows an oscillatory behavior with $L/\lambda_{F}$, with
an amplitude which takes the value 1 for some ranges of the length. We note that
this perfect pseudospin valve effect of the monolayer graphene is
more robust with respect to an increase of the length of the N
region, as compared to the similar effect in a bilayer graphene
pseudospin valve structure\cite{Jose09}.
\par
The amplitude of PMR decreases by increasing $\mu/\Delta$ and
tends to the constant value of $PMR=1/3$ for a highly
doped structure with $\mu\gg\Delta$. This residual PMR is the
difference in the resistance of a n-p graphene structure with that
of a uniformly (p or n) doped graphene with the same $|\mu|$,
which is present even in the limit $PM\rightarrow 0$ of a
non-pseudomagnetized structure.
\par
\begin{figure}
\begin{center}
\includegraphics[width=3in]{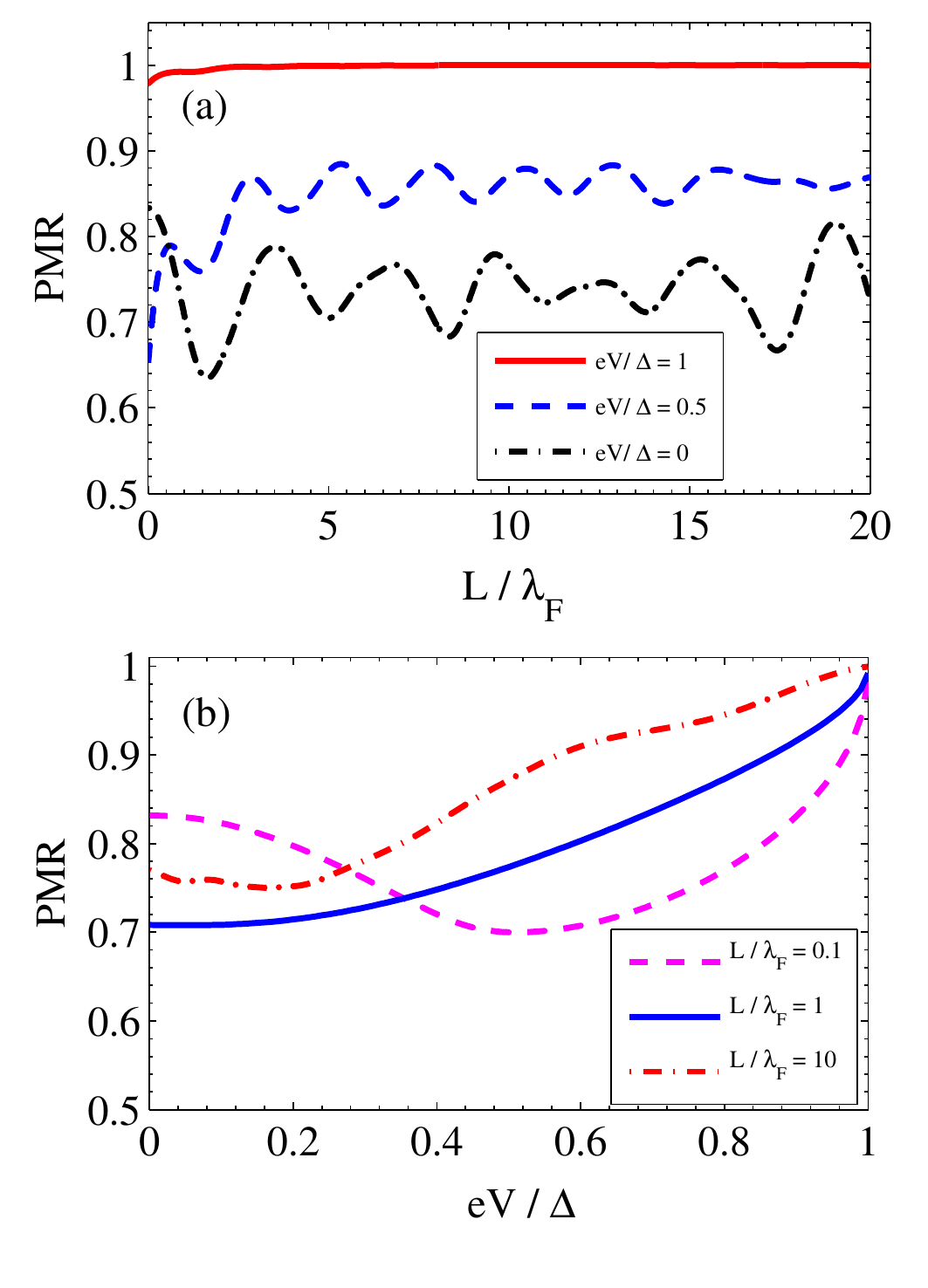}
\end{center}
\caption{\label{Fig:3}(Color online)(a) Plot of PMR versus $L/\lambda_F$ for three values of bias voltage
$eV/\Delta=0,0.5,1$ (b) and the behavior of PMR versus $eV/\Delta$ for three values of $L/\lambda_F$, when $\mu=2\ \Delta$.}
\end{figure}
We have found that the perfect pseudospin valve effect can be
resumed by applying a bias voltage $V$ of an appropriate amplitude
to the valve. This is shown in Fig. \ref{Fig:3}a,b. In Fig. \ref{Fig:3}a, PMR is
plotted versus $L/\lambda_F$ for $\mu/\Delta = 2$ and different
$eV/\Delta$. As it is seen, by increasing eV from 0 the amplitude
of PMR increases and the oscillatory behaviour is suppressed, such
that PMR reaches the perfect constant value $1$  when $eV=\Delta$.
Fig. \ref{Fig:3}b shows plot of PMR as a function of $eV/\Delta$
for different values of $L/\lambda_F$, which shows that the
pseudospin valve becomes perfect as $eV\rightarrow1$, independent
of the value of $L/\lambda_F$. Applying bias voltage is such that there is no band crossing for p-doped
excitations ($eV\leq\mu-\Delta$). This is similar to the case of
having PF regions with different chemical potentials, where
$\mu_{n,p} = eV\pm \mu$ and $\mu_n\geq|\mu_p|$.

\section{\label{sec:level2}Proximity effect in PF/N Junctions}

The above found strong robustness of the pseudospin valve effect
with respect to increasing the length of N contact should arise
from a strong pseudomagnetic coupling between the two graphene PF
regions. This itself can be due to a long range penetration of
pseudospin polarization into the N region by proximity to PF
regions. To analyze this in detail, we study proximity effect in
PF/N, PF/N/PF junctions.  We start with a single PF/N junction in
a graphene sheet in the x-y plane, where the region $x<0$ (PF
region) has a uniform PM oriented vertically to the sheet and the
region $x>0$ (N region) is in the normal state. We calculate PM in
PF and N regions using Eq. (\ref{pseudomagnetization}), and by
considering the contribution of the pseudospin of all incident
electrons from left and right regions that are scattered from the
junction. The total pseudomagnetization vector $\bm{PM}$ will be
the sum of PM vectors contributed from electrons incident from the
left and the right regions:
\begin{eqnarray}
\label{PM_proximity}
&&\frac{\bm{PM}_j}{N_j}=\frac{1}{2}\{\frac{\bm{PM}_j^{l}}{N_j^{l}}+\frac{\bm{PM}_j^{r}}{N_j^{r}}\},\nonumber\\
&&\bm{PM}_j^{l(r)}=\sum_{\bm{k}}<\bm{\sigma}(\bm{k})>_{\psi_{j,l(r)}},\\
&&N_j^{l(r)}=\sum_{\bm{k}}\psi_{j,l(r)}^*\psi_{j,l(r)},\nonumber
\end{eqnarray}
where $j$ denotes the PF(N) region.
\begin{figure}
\begin{center}
\includegraphics[width=3in]{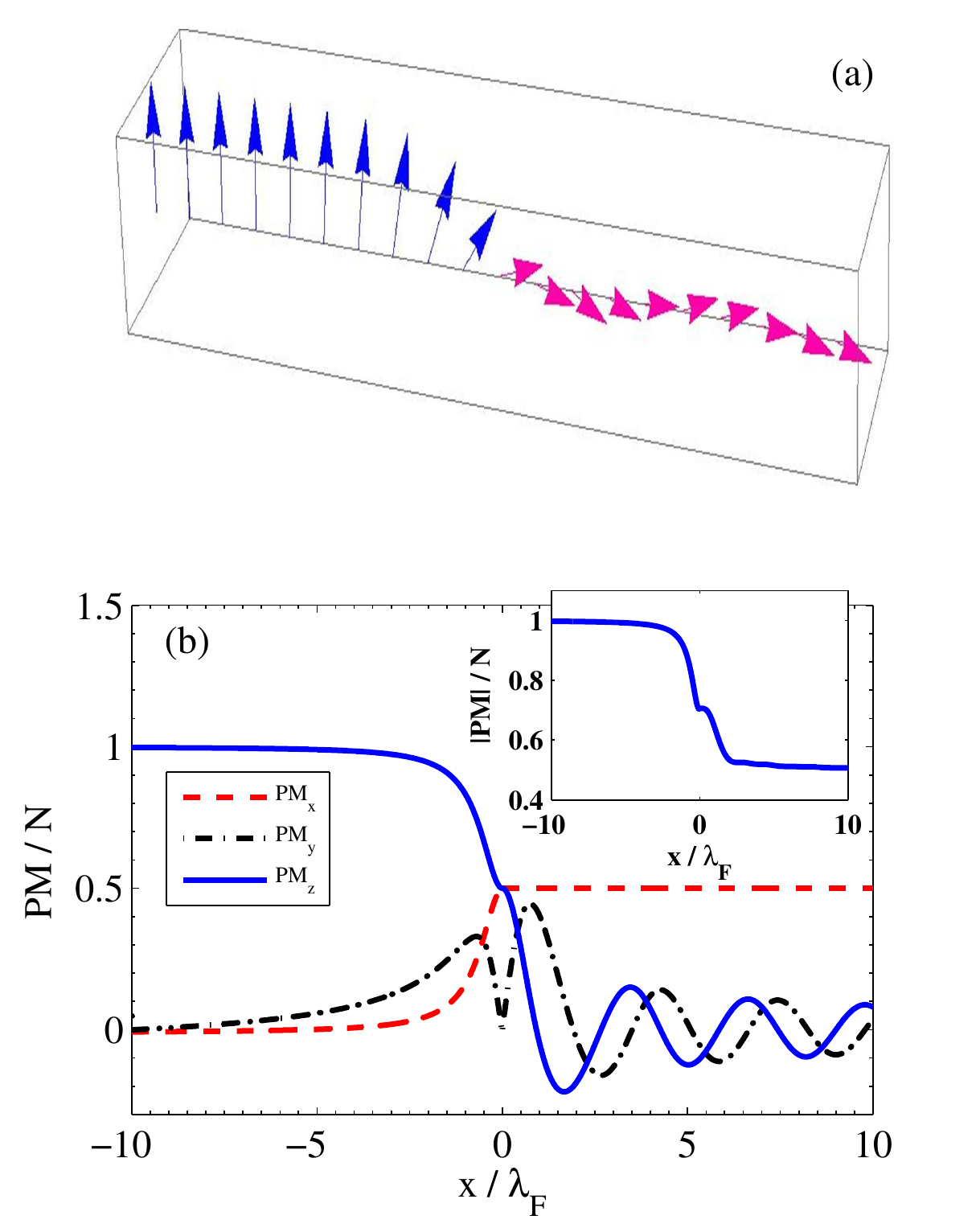}
\end{center}
\caption{\label{Fig:4}(Color online) PM of the PF/N junction when $\mu\simeq\Delta$:
The profile of $\bm{PM}$ in PF (blue) and N regions (pink) (a) and the position dependence of the $\bm{PM}$ components (b).
The inset of Fig. \ref{Fig:4}b shows the magnitude of $\bm{PM}$. }
\end{figure}
\par
The resulting profile of $\bm{PM}$ across the PF/N
junction is demonstrated in Fig. \ref{Fig:4} for
$\mu\simeq\Delta$. As it is seen, a nonzero $\bm{PM}$ is induced
in N region ($\Delta=0$) which rotates around the normal to the
junction (x axis) with $x$. The perpendicular component $PM_z$
oscillates as a function of $x$ with a period of order
$\lambda_F$, and shows only a weak decay in the scale of
$\lambda_F$. While the in-plane components $PM_{x,y}$ vanish inside
 the PF regions, they are produced at the PF/N interface and
are penetrated into the N region. This is originated from  the
difference between the pseudospin of electrons coming from PF side
(Eq. (\ref{pseudospin}) ) and that of electrons coming from N side (Eq. (\ref{pseudospin}) with
$\Delta=0$), which results in a nonvanishing averaged over
momentum directions pseudospin polarization.  The $y$ component
$PM_{y}$ shows an oscillatory behaviour with $x$ similar to
$PM_{z}$. Interestingly, $PM_x$ is uniform inside N, which
considering the decay of the other two components, implies that
$\bm{PM}$ at the points in N far from the junction is uniform and
oriented perpendicular (along x axis) to the $\bm{PM}$ in the connected
PF. This unusual proximity effect can be explained in terms of
reflectionless Klein transmission of electrons which incident
normally to PF/N interface\cite{Katsnelson06,Young09,Stander09}. We note to the unusually long
range penetration of the proximity induced PM inside the N region, which is in contrast to the ferromagnet/normal metal
junction (FN), in which the induced magnetization decays over
short interatomic distances.
\par
The above analysis of the proximity effect in PF/N junction can be
extended to the pseudospin valve geometry of Fig. \ref{Fig:1}b.
 The profile of $\bm{PM}$ orientation
in different regions of the PF/N/PF junction is indicated in Fig.
\ref{Fig:1} for parallel (c) and antiparallel (d) cases when
$L=\lambda_F$ and $\mu\simeq\Delta$. $\bm{PM}$ is
perpendicular to the x-axis and undergoes rotation across the N
contact in a way that in P and AP cases $PM_y$ and $PM_z$,
respectively, shows change of signs at the middle of N region
($x=L/2$).
\par
\begin{figure}
\begin{center}
\includegraphics[width=3in]{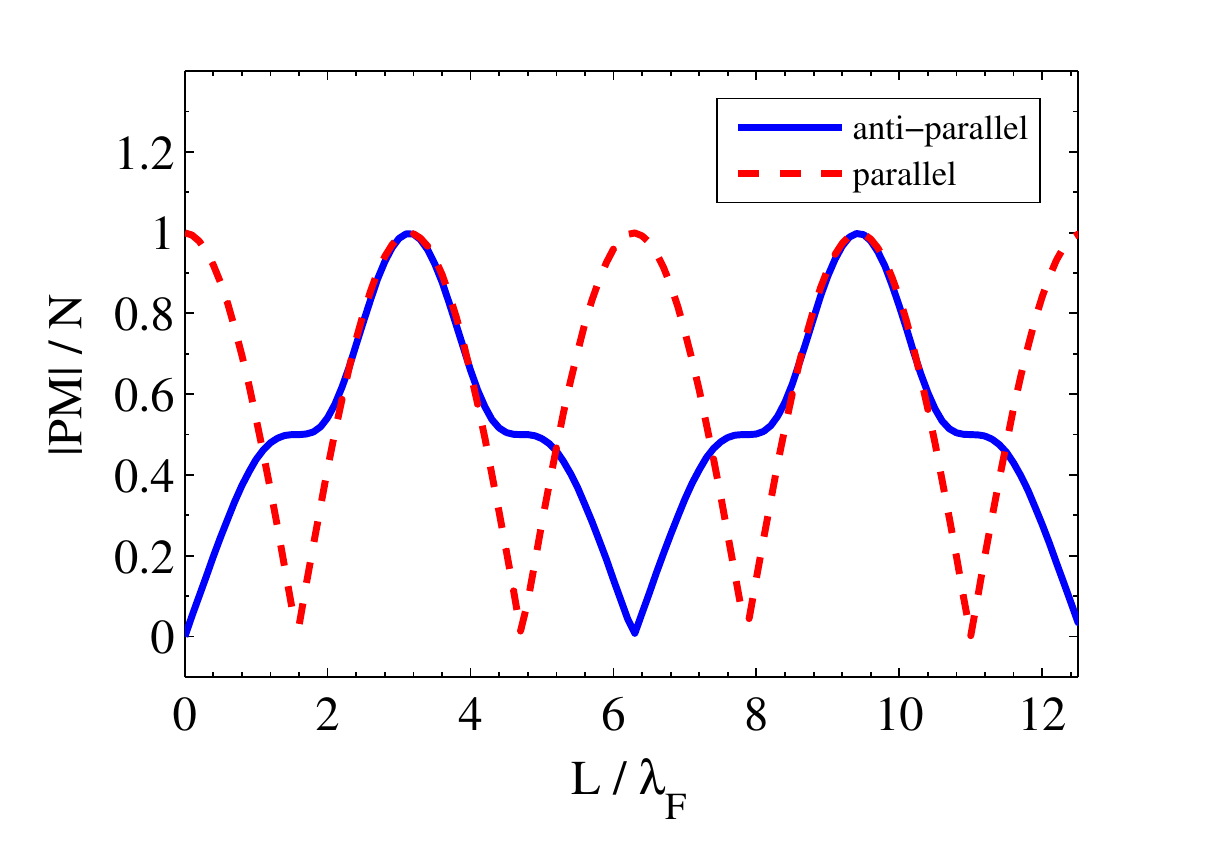}
\end{center}
\caption{\label{Fig:5}(Color online) Plot of PM dependance on the length of the N region
$L/\lambda_F$ for PF/N/PF junctions, where $\mu\simeq\Delta$. }
\end{figure}

Furthermore, we obtain that the magnitude of $\bm{PM}$ inside the
N region is constant with x for both of parallel and
anti-parallel configurations.  Dependence of the magnitude of
$\bm{PM}$ on the length of the N region L and for $\mu/\Delta
\simeq 1$, is shown in Fig .\ref{Fig:5} for both of parallel and
anti-parallel configurations. We see that they have a periodic
behavior with L. As is expected, in the limit $L\rightarrow0$
pseudomagnetization of the parallel configuration goes to the
constant value of a PF layer $|PM|/N=1$, while it
tends to zero for anti-parallel configuration (n-p junction).
\par

\section{\label{sec:level3}Conclusion}

In conclusion, we have investigated realization of pseudospin
polarized quantum transport in monolayer graphene with an energy
gap in its Dirac spectrum. We have demonstrated that a gapped
graphene is a pseudospin symmetry-broken state which exhibits a
pseudomagnetization PM oriented normal to plane of the graphene. The
magnitude of PM depends on the ratio of the
chemical potential to the energy gap $\mu/\Delta$ and its direction
is switched by changing the type of doping between n and p. Based on
this observation, we have proposed a pseudospin valve in which a non-pseudomagnetized
normal region connects two pseudoferromagnetic (PF) regions, whose
magnetization alignment can be controlled by altering the type of
their doping. The suggested pseudospin valve exhibits a pseudomagnetoresistance PMR,
 defined as the relative difference of the
resistances in parallel and antiparallel alignments, which for
$\mu\simeq\Delta$ can be tuned to unity by appropriately adjusting
the contact length $L$. We have shown that this perfect
pseudomagnetic valve effect with $PMR=1$ is preserved even for
very large lengths $L\gg \lambda_F$. Although PMR decreases by
increasing $\mu/\Delta$ in the absence of a bias voltage, a perfect
switching at large chemical potentials can be resumed by
applying an appropriate bias voltage.
\par
In order to explain the robustness of the pseudospin valve effect
with respect to increasing of the contact length, we have further
studied the proximity effect in PF/N junctions. We have
found that an equilibrium pseudospin polarization can be induced
in the normal graphene over a very large length $L\gg \lambda_F$.
The induced pseudomagnetization vector $\bm{PM}$ undergoes a
damped spatial precession  around the normal to the PF/N junction
and tends to be uniform along the normal at the large distances
$x\gg \lambda_F$ from the junction.

\begin{acknowledgments}
We thank A. G. Moghaddam for fruitful discussions. The authors
gratefully acknowledge support by the Institute for Advanced Studies
in Basic Sciences (IASBS) Research Council under grant No. G2009IASBS110.
\end{acknowledgments}

\end{document}